\documentclass[12pt]{article}
\usepackage[dvips]{graphicx}
\begin{document}
\date{\today}
\title{TDHF simulation of the expansion of abraded nuclei.}
\author{D.~Lacroix$^1$, Ph. Chomaz$^1$ \\
\medskip $(1)$G.A.N.I.L, B.P. 5027, F-14076 Caen Cedex 5, France.}
\maketitle

\begin{abstract}
A recent interpretation of the caloric curve based on the expansion of the
abraded spectator nucleus is re-analysed in the framework of the 
Time-Dependent Hartree-Fock (TDHF) evolution. It is shown that the TDHF dynamics
is more complex than a single monopolar collective motion at moderate
energy. The inclusion
of other important collective degrees of freedom may lead to the dynamical
creation of hollow structure. Then, low density regions could be locally
reached after a long time by the creation of these exotic density profiles.
In particular the systematic of the minimum density reached during the
expansion (the so-called turning points) appears to be different. 
\end{abstract}

{\bf PACS:} 25.70.-z, 24.10.Cn, 24.10.-i, 25.70.Gh, 24.30.Cz \\

{\bf Keywords:} Mean-Field, TDHF, Compound nuclei, Collective motions.
\newpage
\section{Introduction}

During the past decade, many works both in experimental and theoretical
nuclear physics, have been devoted to the search of the liquid-gas phase
transition in Heavy-Ion collisions at intermediate energies. One of the most
striking results is the observation of the so-called caloric curve \cite
{Poch1}. Indeed, using an isotopic thermometer in conjunction with a measure
of the total excitation energy, the authors of ref. \cite{Poch1} have
observed that, over a wide range of excitation energy, the temperature  
of the abraded source
seems
constant while for a system in an unique phase, one would expect a monotonic
increase of the temperature with excitation energy. This curve is similar to
those known for liquid-gas phase transition at constant pressure. Indeed, in
a coexistence region  
of a 
single fluid system, it exists a univocal relation between all intensive
thermodynamical quantities. From the experimental
point of view this curve is actually subject of debates since neither the
temperature nor the excitation energy are well defined quantities for small
objects in rapid evolution such as hot nuclei\cite{Indra}.

Despite all these difficulties, this observation might be a major step
forward since it could be a direct evidence of the liquid-gas phase
transition in nuclear matter \cite{Calor}. Recently, an interpretation of
this curve has been proposed by Papp and N\"{o}renberg\cite{Papp1}.
Considering the expansion of excited projectile like fragments in a
collective model, they predict that the temperatures at maximum dilution
follow approximately the experimental caloric curve. 
Indeed, the temperature remains almost constant while 
more and more excited systems are considered because the densities at
the turning point of the monopole oscillation rapidly
decrease as a the initial excitation increase. 
As stressed in a recent review article\cite{Poch97}, 
this interpretation
is of great importance since,  
it gives an intuitive connection between
the creation of compound nucleus, radial flow, multifragmentation and
the final results of Heavy-Ion collisions.  

In this paper we would
like to re-investigate the same scenario within a microscopic approach:
the time-dependent Hartree-Fock (TDHF) dynamics. 
This is not the first attempt in this direction, 
since TDHF\cite{Vau1,evap,Strack} as well as BUU\cite{bubble,BUU} have 
been used in the past to study expanding nuclear systems. However,
in this article, we will concentrate on the possible differences 
between a full quantum treatment and a macroscopic collective 
model\cite{Papp1}.

In the next chapter, we will recall the general framework of the work done
in ref.\cite{Papp1}. In part $3$, we describe features of the microscopic
model we have used. Two forces of Skyrme type are used, 
one leads to a Soft Equation of
State (EOS) in the infinite nuclear matter, the other one to a stiff EOS.
These EOS are key issues when discussing liquid-gas phase transition and in
particular mechanical instabilities occurring inside the coexistence zone,
namely spinodale instabilities. In this article, we concentrate on the
dynamics of expanding systems as predicted by the time-dependent
Hartree-Fock theory. In order to discuss the observed results we have defined
an equivalent of the EOS for finite systems. We have pointed out
many differences between the energy of finite systems as a function of the
dilution and the infinite nuclear matter case, this 
could be in part understood in terms of 
surface effect and Coulomb interaction. 

We have paid a particular attention
in initial conditions, in order to be in same framework as in ref. \cite
{Papp1}. The authors of ref. \cite{Papp1} follows the idea of a
dominant monopolar collective mode. This idea was used many times in
Heavy-Ion Physics at intermediate energy in order to have quantitative
informations on how an excited expanding uniform sphere could break\cite
{simple}. Our calculation do not make this assumption, since we do not
impose the domination of one collective mode on the others\cite{Ring}. In
this more general framework, the expansion dynamics appears more complex
than the one described in ref. \cite{Papp1}. In
particular, 
for moderate excitation energies (T $<$ 5 MeV) 
we show that hollow structures can be formed during the TDHF
dynamics after several monopolar oscillations. These exotic shapes
are not included in breathing mode pictures\cite{simple} and
could change our physical understanding of Heavy-Ion collisions
at intermediate energy. In a recent work\cite{breathing}, we have shown,
that other collective degrees of freedom should be included in order to
describe the TDHF expansion. In the following, we will generalize the simple
breathing mode picture in order to take into account more collective
variables. In particular, we illustrate the inclusion of more accessible
density profile by defining a generalized EOS that include hollow density
profiles.
On the other hand, considering highly excited sources (T $>$ 5 MeV) we
observe that the quantum dynamics differs from collective models because
of the importance of wave propagation and of thermal mixing 
on the dynamics: barrier transmision,
large amplitude motion, lost of collectivity, ...

\smallskip

\section{\protect\smallskip One model for the interpretation of the caloric
curve}

Let us first recall the scenario followed by Papp ad N\"{o}renberg in order
to understand the ALADIN results concerning the fragmentation of a gold
projectile after its interaction with a gold target at 600 MeV/A incident
energy\cite{Poch1,Papp1}. Since the experiment is
observing the decay of the projectile-like fragments in peripheral heavy ion
reactions, the initial conditions are provided by the abrasion-ablation model
\cite{Gaim1}: after collision, the projectile of mass $A$ has lost $\Delta
A_{i}$ nucleons; the initial excitation $E_{i}^{\star }$ of the nucleus is
assumed to be related to $\Delta A_i$ by 
\begin{equation}
E_{i}^{\star }=\gamma \Delta A_{i}  \label{abr}
\end{equation}

\noindent In the following simulations,
$\Delta A_{i}$ varies from 8 to 108 nucleons and $\gamma
=13.3MeV$. Temperatures are adjusted in order to get the excitation
energies $E_{i}^{\star }$ (\ref{abr}) for a spherical nucleus of mass $A$-
$\Delta A_{i}$.
The initial density is taken to be lower than the normal
density, $\rho _{i}=0.8\;\rho _{0}$ because the abrasion-ablation dynamics
is assumed to induce a small
dilution through the ejection of few nucleons\cite{Papp1}.

Then, authors of ref.\cite{Papp1} follow the expansion of the system 
as an isentropic self-similar
monopole motion: the breathing mode. In this picture, 
the internal degrees of freedom are assumed to be
equilibrated like in the hydrodynamical regime.
However, the energy and the entropy 
slightly evolve in time in order to account for particles evaporation.

Studying the evolution of different such systems with various initial
masses, they have shown that, choosing a soft equation of state, the caloric
curve may be explained as the temperature  
of the emitting source 
at the turning points of the
collective expansion associated with various initial masses and excitation
energies. 
They have also
found that using a stiff equation of states reduces the amplitude of the
considered monopole vibration leading to different conclusions.

The aim of our paper, is to examine these conclusions within microscopic
model known to describe collective dynamics as well as single particle
behaviors in a quantum framework: the TDHF\ approach. In particular, we will
test the validity of the breathing mode scenario by comparison with the
complete TDHF dynamics.

\section{\protect\smallskip TDHF\ simulations}

\subsection{\protect\smallskip Mean-field approximation}

We consider the evolution of hot diluted spherical nuclei
in the framework of mean-field theory. The simulation of the desexcitations
of excited compressed spherical nuclei is already reported in the literature%
\cite{Vau1} for nuclei initialized using a Constrained Hartree-Fock method
(CHF).  
We use a scaling assumptions in order to prepare
the initial nucleus because this method allows to study a broad range of
initial dilutions. In the cases where both methods are
tractable, we have controlled that 
it gives results comparable with those of CHF \cite{denis1}. 

We consider systems saturated in spin and isospin.
Nucleons are moving independently in an average spherically-symmetric
potential. We use the following parameterization for the effective
mean-field potential 
\begin{equation}
U\left[ \rho \right] =\frac{3}{4}~t_{0}~\rho +\frac{(\sigma +2)}{16}%
~t_{3}~\rho ^{\sigma +1}+c~\nabla ^{2}\rho +V_{C}.  \label{Skyrme}
\end{equation}
\noindent \noindent \noindent where the surface term, $c~\nabla ^{2}\rho ,$
can be related to the usual Skyrme parameters by

\begin{equation}
c=\frac{5t_{2}-3t_{1}}{16}
\end{equation}

\noindent \noindent For sake of simplicity we have
taken forces with effective masses equal to the bare one. In order to reduce
the numerical instabilities we have used the standard method which consists
in replacing the
surface term of potential (\ref{Skyrme}) by a folding product with a finite
range function \cite{Koo1,bon1} 
\begin{equation}
U\left[ \rho \right] =\frac{3}{4}t_{0}^{\prime }\;\rho +\frac{(\sigma +2)}{16%
}t_{3}\;{\rho }^{\sigma +1}+V_{0}\;Y\otimes {\rho }+V_{C}  \label{pot}
\end{equation}

\smallskip \noindent \noindent \noindent where $Y$ is a Yukawa folding
function 
\begin{equation}
Y(\vec{r})={\frac{\exp \left( -\frac{r}{a}\right) }{\frac{r}{a}}}
\end{equation}

\smallskip 

\noindent at the lowest order in the range $a$ this expression is equivalent
to the usual Skyrme functional provided that

\begin{equation}
t_{0}^{\prime }=t_{0}-\frac{16}{3}\pi a^{3}V_{0}  
\label{T_0}
\end{equation}

\smallskip \noindent \noindent and that coefficients in the Yukawa folding
functions are related by

\begin{equation}
V_{0}=\frac{(5t_{2}-3t_{1})}{64\pi a^{5}}=\frac{c}{4\pi a^{5}}  \label{V_0}
\end{equation}

Finally, the direct Coulomb potential $V_{C}$ is introduced in an
approximative way by giving an effective charge of $+\frac{Z}{A}$ to each
nucleon. Using this potential and a spherically symmetric density, each
single-particle wave function can be separated into its radial, angular and
spin-isospin part 
\begin{equation}
\Phi _{\alpha }(\vec{r},\sigma ,\tau )=\frac{R_{nl}(r)}{r}Y_{lm}(\theta
,\varphi )\chi _{s}(\sigma )\chi _{t}(\tau )  \label{phi}
\end{equation}
\noindent where $\alpha $ represents all quantum number $\alpha =(n,l,m,s,t)$
, in which $n$ is the energy principal number, $(l,m)$ are the 
usual angular-momentum quantum numbers
and $(s,t)$ is the quantization of spin and isospin.
If we consider initially a hot system, at temperature $T$, occupation
numbers of various orbitals $\alpha $ are given by a Fermi-Dirac
distribution 
\begin{equation}
n_{nl}=\frac{1}{\exp \left( \frac{\varepsilon _{nl}-\mu }{T}%
\right) +1}  \label{Fermi}
\end{equation}
\noindent where the chemical potential is computed in order to get the
correct number of particles. To be able
to define particle orbitals at high temperature,
a small external field $\lambda r^{2}$ has
been added to the potential ($\lambda =0.25$ MeV/fm$^{2}$). Finally, the
density of our spherical nucleus takes the particular form

\begin{equation}
\rho (r,t)=4\sum_{n,l}(2l+1)n_{nl} \frac{\left| R_{nl}(r,t)\right|^{2}}{4\pi
r^{2}},  \label{eq:Rho}
\end{equation}

Before getting into details of the 
simulations, we will give some more details 
about these forces and about the associated
Equation Of State (EOS).

\smallskip

\subsection{{\bf Forces and infinite nuclear matter properties}}

Following \cite{Papp1}, we consider two parametrizations of the
potential (\ref{pot}) equivalent to the $SIII$ and the $SkM^{*}$ forces
(\cite{Bra1}). These two forces correspond respectively to a HARD and a SOFT
EOS.

\smallskip

 Starting with the mean-field potential (\ref{Skyrme}), the total energy
per nucleon in the infinite, uniform, spin and isospin saturated nuclear
matter is 
\begin{equation}
\frac{E}{A}=\frac{E_{k}}{A}+\frac{3}{8}t_{0}\rho _{0}+\frac{1}{16}t_{3}
\rho_{0}^{(\sigma +1)}
\label{EOS1}  
\end{equation}
\noindent \noindent where we have introduced the kinetic energy per nucleons 
$\frac{E_{k}}{A}$; for a cold gas of fermions this energy can be easily
related to the density by 
\begin{equation}
\frac{E_{k}}{A}=\frac{3}{5}\left( \frac{\hbar ^{2}}{2m}\right) \left( \frac{%
3\pi ^{2}}{2}\rho _{0}\right) ^{\frac{2}{3}}  \label{eq:EKIN}
\end{equation}
\noindent Therefore, the saturation density, $\rho _{0}$, is solution of 
\begin{equation}
\rho _{0}\left( \frac{\partial \frac{E}{A}}{\partial \rho }\right) _{\rho
=\rho _{0}}=0=\frac{2}{3}\frac{E_{k}}{A}+\frac{3}{8}t_{0}\rho _{0}+\frac{%
(\sigma +1)}{16}t_{3}\rho _{0}^{(\sigma +1)}  \label{EOS2}
\end{equation}
\noindent which is nothing but the annulation of the pressure at the
saturation point (at zero temperature). The infinite nuclear matter
incompressibility $K_{\infty }\ $is then given by 
\begin{equation}
K_{\infty }=9\rho _{0}^{2}\left( \frac{\partial ^{2}\frac{E}{A}}{\partial
\rho ^{2}}\right) _{\rho =\rho _{0}}=-2\frac{E_{k}}{A}+\frac{9\sigma (\sigma
+1)}{16}t_{3}\rho _{0}^{(1+\sigma )}  \label{EOS3}
\end{equation}
These three relations (\ref{EOS1}), (\ref{EOS2}) and (\ref{EOS3}) provide an
univocal relation between the saturation point properties $\rho _{0}$, $%
\frac{E}{A}$ and $K_{\infty }$ and the force parameters $t_{0},t_{3}$ and $%
\sigma $. The parameter $c$ does not influence the infinite uniform
medium properties.

If we compute the EOS associated with the parametrization (\ref{pot}) we get
similar expressions provided that we use the relation (\ref{T_0}): $%
t_{0}=t_{0}^{\prime }+\frac{16}{3}\pi a^{3}V_{0}$. The parameter $V_{0}$ and 
$a$ are related through the relation (\ref{V_0}) but are not directly
constrained by the EOS. However, for a finite
system the relation between (\ref{Skyrme}) and (\ref{pot}) is only valid at
the lowest order in $a$. Therefore, this parameter should remains small.
Following ref. \cite{bon1} we have fixed $a$ to $0.45979fm.$ Parameters
values leading to same properties as $SIII$ and $SkM^{*}$ for our
parameterization are shown in table \ref{table1}.

\smallskip

As an illustration, we have plotted the various isentropic EOS in 
infinite nuclear matter in figure \ref
{fig1_EOS} (Top). We have also plotted the isentropic and the isothermal
spinodal regions defined as the region where the
derivative of the pressure versus the density at constant entropy
or temperature is negative. Therefore, in
these regions the matter is mechanically unstable against density
fluctuations: these are the so-called spinodal instabilities which might be
responsible for the fragmentation of the system.

\smallskip

Let us now come to the simulation of finite systems. These numerical
simulations consist in two distinct parts: the first one is the
initialization of abraded nuclei and the second one the TDHF dynamics.

\smallskip

\subsection{\protect\smallskip Initialization of finite systems}

The first step is to initialize our system according to the equation (\ref
{abr}). The method used in order to generate a hot and diluted initial
source proceeds as follow:

\begin{itemize}
\item  First we look for the ground state of the nucleus with $%
A_{i}=A-\Delta A_{i}$ nucleons. To do so we use the imaginary time method 
with a small constraint $\lambda r^2$ added to the mean field in
order to get occupied states, $\left| \Phi _{h}\right\rangle $ ( $h$ for
hole states), and the associated one-body density $\rho =\sum_{h}\left| \Phi
_{h}\right\rangle n_{h}\left\langle \Phi _{h}\right| $ where the occupation
number are given by equation (\ref{Fermi}) at zero temperature. 
Diagonalizing the Schr\"{o}dinger
equation 
\begin{equation}
\left\{ \frac{-\hbar ^{2}}{2m}\frac{d}{d{r^{2}}}+\frac{\hbar ^{2}l(l+1)}{%
2mr^{2}}+U(\rho (r))+\lambda r^{2}\right\} R_{nl}(r)=\varepsilon
_{nl}R_{nl}(r)
\end{equation}
we can define the single particle states $\left( nl\right) $. 
The small constraint $\lambda r^{2}$ insures a limited 
number of orbitals for treating the continuum. 
In analogy with the nuclear matter, we define the
saturation density, $\rho _{0}\left( A_{i}\right) $, as the central averaged
density in a sphere of $2fm$.

\item  Then, single particle states are occupied 
according to a Fermi-Dirac statistic (%
\ref{Fermi}) \noindent using a temperature $T$ as a free parameter. This
temperature is defined iteratively in such a way that after the scaling
described below, we get the excitation energy defined by Eq. (\ref{abr})%
\footnote{%
For each $\Delta A_{i}$, we can define the excitation energy by subtracting
the ground state energy of a cold nucleus of size $A-\Delta A_{i} $ to the
energy of the excited nucleus after rescaling. For all these calculations, we
have used the expression of the energy corresponding to the potential (\ref
{pot}) 
\begin{equation}
E=E_{kin}+\int \left\{ \frac{3}{8}t_{0}\rho (\vec{r})^{2}+\frac{1}{16}
t_{3}\rho (\vec{r})^{(\sigma +2)}+\frac{V_{0}}{2}\int {Y}\left( r-r^{\prime
}\right) {\rho (r)\rho (r^{\prime })}d\vec{r^{\prime }}\right\} d\vec{r}
+E_{C}
\end{equation}
\noindent \noindent In this expression, $E_{kin}$ and $E_{C}$ are
respectively the kinetic and Coulomb energy.}. The chemical potential is
determined requiring that the total number of particle is $A-\Delta A_{i}.$
The initial entropy of the system is defined by 
\begin{equation}
S=-4\sum_{nl}\left( 2l+1\right) \left( {n_{nl}\log {n_{nl}}+(1-n_{nl})\log
(1-n_{nl})}\right)  \label{entropy}
\end{equation}

\item  Finally we apply a scaling to each radial wave functions 
\begin{equation}
R_{nl}^{new}(r)={\cal N}R_{nl}(\frac{r}{\delta })  \label{scaling}
\end{equation}

\noindent where ${\cal N}$ is a normalization factor. Then we compute the
average density $\rho \left( A_{i}\right) $ at the center of the nucleus
(within a sphere of 2 fm) and we fix $\delta $ in order to get the desired 
dilatation factor $%
\eta =\frac{\rho \left( A_{i}\right) }{\rho _{0}\left( A_{i}\right) }=0.8$.
\end{itemize}

After the initialization step we are following an ensemble of diluted
nuclei with mass $A-\Delta A_{i}$ and corresponding 
excitation energy $E_{i}^{\star }$. 
Each one is associated with an initial entropy $S_{i}$. A sample of
various initial conditions are given in table \ref{table2}.
Figure \ref{fig2_INI} presents various density profiles. On this graph one
can spot only minor differences between the two forces.

\begin{figure}[tbph]
\begin{center}
\includegraphics*[height=15cm,width=10cm]{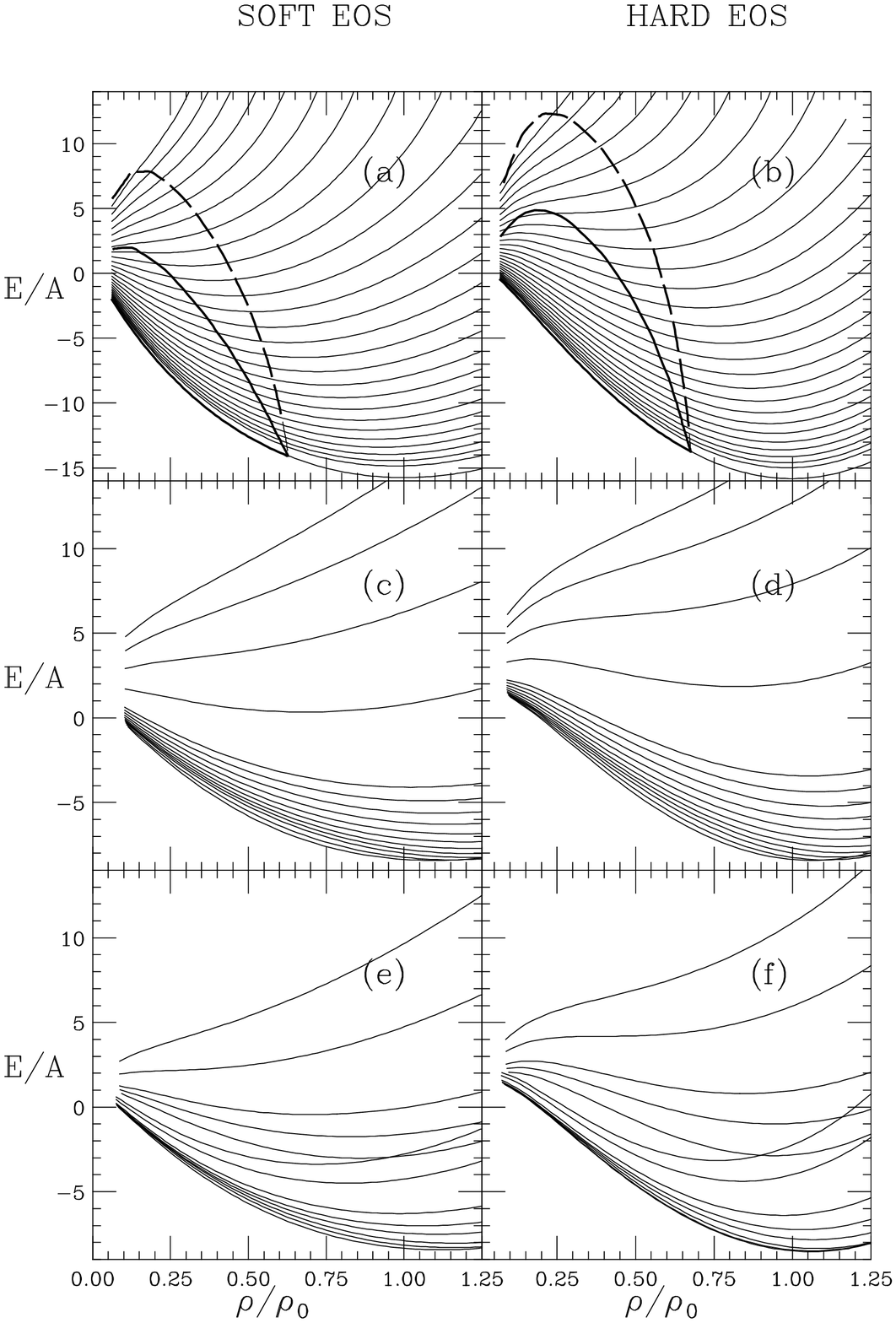}
\end{center}
\caption{Relations between the energy per nucleon and the central
density. Top parts correspond to the infinit medium EOS, 
each curve is drawn at constant entropy and the entropy
variation between two adjacent curves is $\Delta S/k_B=1$. In such a case both
the isothermal (dashed) and the isentropic (solid) spinodal are displayed as
thick lines. Middle part corresponds to a $_{79}^{197}$Au nucleus while the
bottom part is associated with the various systems considered for the
evolution. In the bottom part, curves correspond to decreasing masses
(equivalently to increasing excitation energy) from bottom to top of the
graphic. More detail is given in the text for the determination of this
picture in finite nuclei. For all graphs, left and right part correspond
respectively to a soft and a hard EOS.}
\label{fig1_EOS}
\end{figure}

\begin{figure}[tbph]
\begin{center}
\includegraphics*[height=15cm,width=10cm]{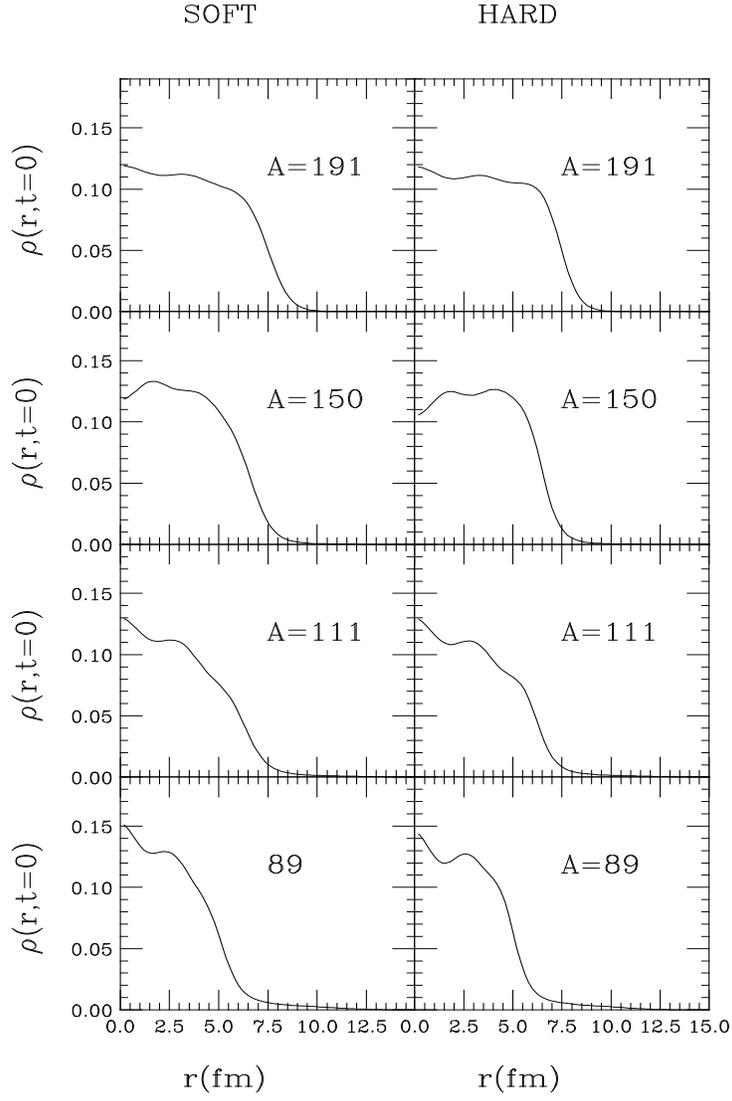}
\end{center}
\caption{Typical examples of initial density profiles for various initial
masses: Left (soft EOS) and right (Hard EOS). }
\label{fig2_INI}
\end{figure}

In order to get a deeper insight in the properties of finite systems, we
have plotted the energy of the considered nuclei as a function of
their central density: $E=f(\eta )$ (fig.(\ref{fig1_EOS}) middle and
bottom). Each curve is obtained by scaling the wave-functions at fixed
occupation numbers (which means that along these curves $S=cte$\footnote{%
These relations between $E$ , $\eta $ and $S$ can
be considered as a finite system equation of states.}). Middle part of fig. (%
\ref{fig1_EOS}) corresponds to a $_{79}^{197}$Au for various entropies
whereas bottom corresponds to masses and entropies reported in table \ref
{table2}.

These curves can be compared to the infinite medium EOS shown in figure \ref
{fig1_EOS} (Top). As far as the dynamics is concerned, these
isentropic curves have only a meaning if the time dependent density profiles
could be obtained one from another by a simple scaling in r-space. 
This is equivalent to consider the breathing mode as dominant and to
neglect other fluctuations in density.

Figure 1 illustrates 
the fact that finite systems have
surface and Coulomb contributions to their energy. This modifies the
saturation energy to 8-9 MeV (depending upon the mass of the nucleus
considered) instead of 16 MeV found in the case of a neutral infinite
system. Moreover, it changes the curvature of the isentropes i.e. it changes
the effective compressibility modulus. However, this compressibility 
remains the main
difference between the two ensembles of calculations associated with the two
different forces. These isentropes will help us in the understanding of the
dynamical evolutions since, in a naive picture, the collective vibration
should oscillate around the minimum of the energy for the considered entropy
and the turning points of the breathing mode should be simply obtained by 
requiring energy conservation 
\footnote{ Note that, picture (\ref{fig1_EOS}) depends upon the size of the
sphere we have taken in order to define the density ( here r=2 fm). This
will be discussed in the following.}.

\subsection{\bf TDHF evolution}

Having defined the initial conditions as a given nucleus of size $%
A_{i}=A-\Delta A_{i}$, a given excitation energy and the dilatation
coefficient $\eta =0.8$, we let the system evolve through the TDHF equations 
\footnote{%
For the evolution, we have taken a step of time $\Delta t=0.75fm/c$. We have
discretize the r-space in step of size $\Delta r=0.2fm$, the total size of
the r-space being 300 fm, this size is big enough to avoid bouncing
of evaporated particles against boundaries.}

\begin{center}
\begin{equation}
i\hbar \frac{\partial R_{nl}(r,t)}{\partial {t}}=\left\{ \frac{-\hbar ^{2}}{%
2m}\frac{\partial }{\partial {r^{2}}}+\frac{\hbar ^{2}l(l+1)}{2mr^{2}}%
+U\left[ \rho \right] (r,t)\right\} R_{nl}(r,t)
\end{equation}
\end{center}

The potential term $U\left[ \rho \right] $ is the same as in equation (\ref
{pot}). The TDHF equations have the particularity to conserve occupation
numbers ($n_{nl}$) of single particle levels. 
\begin{equation}
\frac{d}{dt}{n_{nl}}=0
\end{equation}
The entropy of the whole system (\ref{entropy}) is thus conserved during the
evolution. On the other hand the total energy of the system is also
conserved. However, during the evolution, nuclei are evaporating particles
and a tail at large distance will develop for the various wave functions $%
R_{nl}$. We can try to investigate this process using the methods developed
in \cite{evap} which consist in splitting the r-space into two parts :

\begin{itemize}
\item  a sphere containing the initial nucleus, here we have considered a
sphere of radius 15 fm. This part of the system will be called the nucleus
in the rest of this article.

\item  the rest of the space which will be considered as evaporated particles%
\footnote{{\footnotesize We are considering very
large space in order to avoid the reflections of the wave packets on the
boundaries of the considered r-space.}}.
\end{itemize}

\noindent Therefore, due to the particle evaporation, the nucleus is
continuously loosing mass, energy and entropy \footnote{
In general at very low
excitation energy the nucleus can even gain entropy because of the
dissipation of the collective motion.}. 
We are thus in the same framework as in ref.\cite{Papp1,fri1} 
but with a microscopic model.
However, the TDHF approach is neglecting the effect of the collisions 
and this might induce differences between our simulations and those of 
ref.\cite{Papp1,fri1}. These differences are expected to be not too important
because in a self-similar expansion of infinite nuclear matter,
the gain and the loss term due to collisions cancel, so that
the mean field dominates the dynamics. In a finite system, the collisions 
do not exactly cancel as it is known from the calculation of the small
amplitude motion\cite{Sakir}. This should be kept in mind when the various
models are compared.

\begin{center}
\smallskip
\end{center}

\section{Discussion of the expansion dynamics}

\subsection{\protect\smallskip Evolution of global quantities: evaporation
and monopole expansion}

In this section, we study the evolution of the various considered nuclei. In 
fig.(\ref{fig3_EAS}), we have plotted the evolution of masses and entropies of the
systems reported in table \ref{table2}. All these quantities are 
computed within a sphere of 15 fm. From this picture we can see that the
smaller are the initial masses, the larger are the energies and entropies
and so are the number of evaporated particles. Indeed, one can observe that
the evaporation is stronger for light highly excited systems 
with a rapid decrease of the
mass and the entropy. Moreover, in TDHF approach the evaporation is
treated in quantum mechanics, particle may be reflected
several time by the potential well before being ejected 
from the nucleus. Therefore, the particle evaporation appears rather slow.

\noindent This monopole motion can be clearly seen in figure \ref{fig4_rho}
which displays the time evolution of the central densities for different
initial conditions. From this figure, as expected, collective motion
is slower and 
presents a larger amplitude motion in the case of a soft EOS.

\subsection{\protect\smallskip Discussion of the Collective dynamics}

In these figures, we do not observe the fast expansion towards low density
regions. Indeed, at moderate energies
($T<5$ MeV), we
observe that the first minimum (i.e. the first turning point) appears to be
close from the initial dilution (see for example $A=191$). At low
excitation energies, the small dilution of the abrasion-ablation stage is
followed by a recompression phase. Then, when the system expands again,
the average density may reach values
lower than the initial one. 

Conversely, in a simple description  
assuming a collective dynamics at almost constant entropy and mass, 
this behavior should be
normally understood considering the energetics of isentropes as shown in
figure \ref{fig1_EOS}. Indeed, schematically, if the system is initially
with an excitation energy, $E_{0}$, and an entropy, $S_{0}$, it will expand
or contract till it has reached the other point of the isentrope $S=S_{0}$
which cut a horizontal line corresponding to the energy $E=E_{0}$. This
schematic procedure is illustrated in figure \ref{fig7_sch}.
Values of turning points obtained with this simple 
construction as well as the one of ref\cite{Papp1} are compared to those
extracted from the simulation in table \ref{table4}.

\smallskip

\begin{figure}[tbph]
\begin{center}
\includegraphics*[height=15cm,width=12cm]{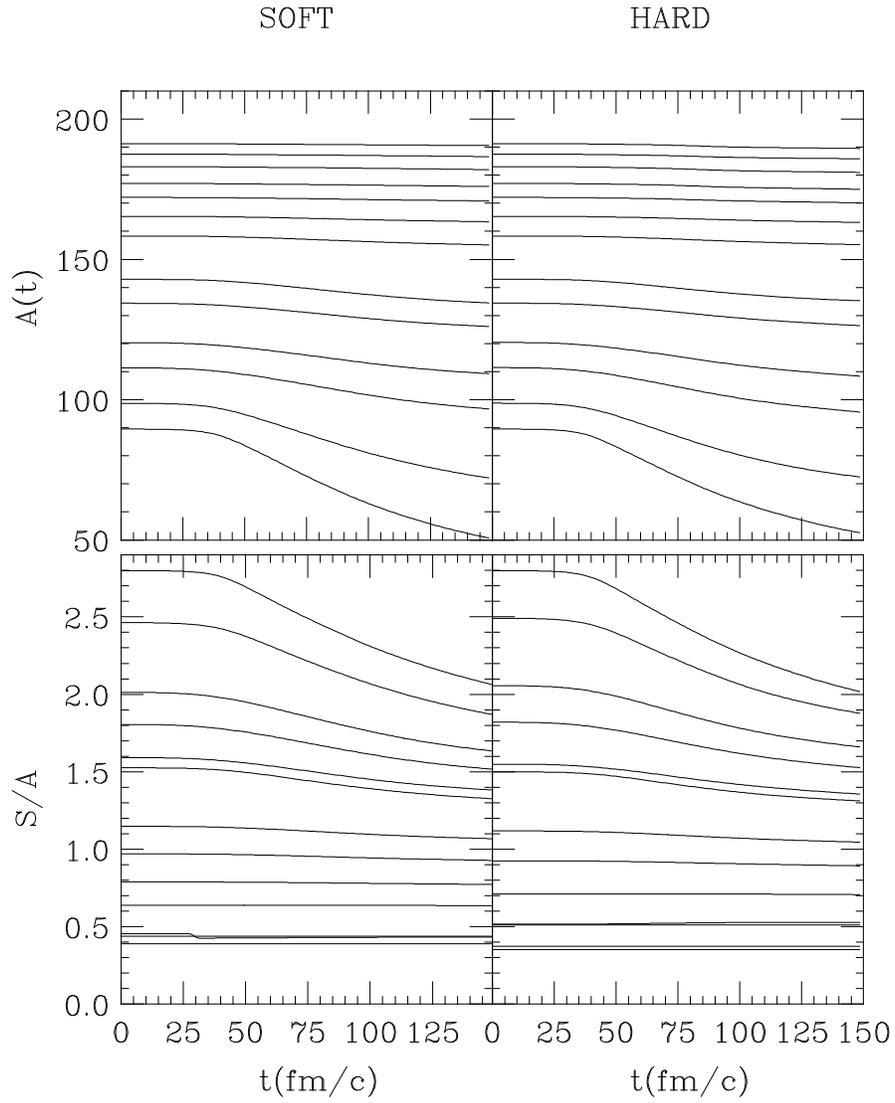}
\end{center}
\caption{Typical examples of time evolution of the mass, and entropy for
the various initial masses considered.}
\label{fig3_EAS}
\end{figure}

\begin{figure}[tbph]
\begin{center}
\includegraphics*[height=15cm,width=10cm]{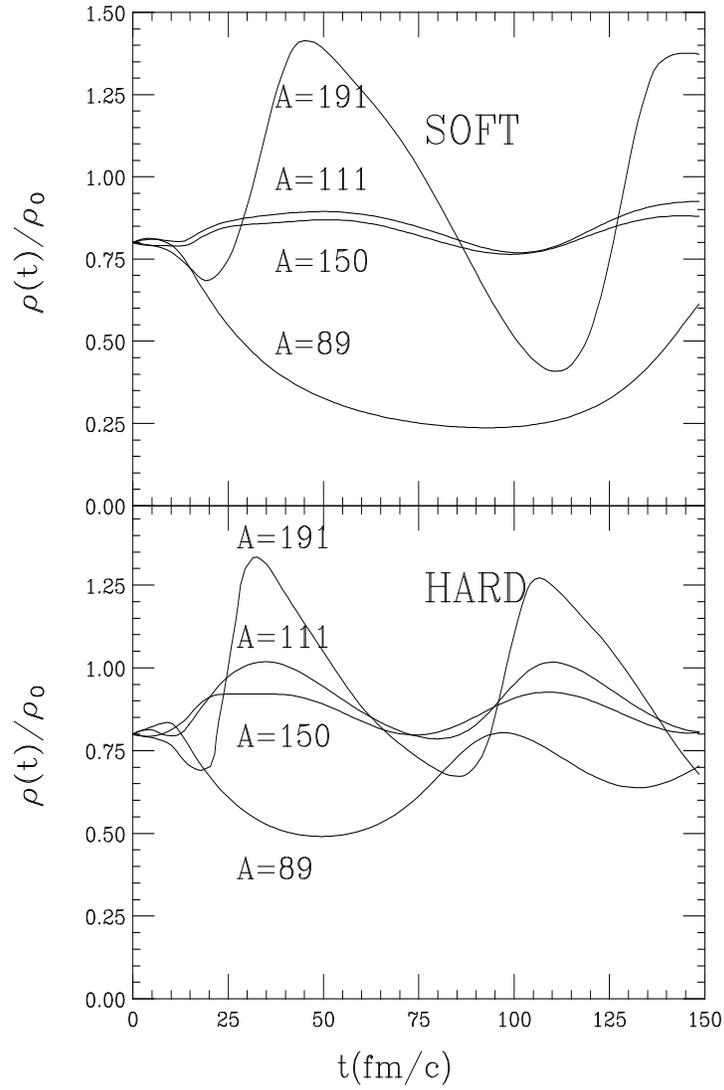}
\end{center}
\caption{Typical examples of the time evolution of the central density for
various initial masses ($A=191, 150, 111$ and $89.4$).}
\label{fig4_rho}
\end{figure}

For $A=150$ and $A=111$, one can see only little differences between
results of the schematic construction and those of the exact dynamics.
However, they are different for $A=191$ and $A=89$.
Considering table \ref{table4} and figure \ref{fig4_rho}, several comments
should be made:

\begin{itemize}
\item  i) For high excitation energies (small masses), the schematic
construction based on energy and entropy conservation predicts a total
vaporization of the nucleus whereas in the dynamical simulations a small
residue survives. This discrepancy can be accounted for by
considering the energy and entropy variation due to the evaporation of
particles  
(see fig.\ref{fig3_EAS}).

\item  ii) for low excitation energy (for example $A=191$), evaporation is
negligeable (S=cte, T=cte (fig. \ref{fig3_EAS})) and the result
could be directly compared to the schematic construction. 
However, we observe that TDHF predictions differ from
schematic predictions. In the schematic model, from figure \ref{fig7_sch}, 
starting with a nucleus of mass $A=191$ at $\eta =0.8$, it appears
impossible to reach a dilution lower than the initial one since the diluted
nucleus is predicted to undergo a compression right after the beginning of
the calculation. In TDHF simulations, the system is
able to reach lower densities than the initial one by creating a hole at the
center of the nucleus. Moreover, instead of undergoing damped monopolar
oscillations, the nucleus reaches a
much lower value of central density at the second dilatation\footnote{
Note that, this effect is not due to the Coulomb interaction  
and so is not analogous to the phenomenon reported by 
Borderie
{\it et} al 
in ref.
\cite{bubble}, 
since removing the Coulomb potential  
from our simulations 
do not 
modify the observed 
behavior. 
}.
\end{itemize}

As far as the comparison of TDHF simulation and the results of \cite{Papp1}
is concerned, the values of the turning points reported
in table \ref{table4} appears to be different.
At high initial temperature ($T>5$ MeV) and for masses lower than $150$,
the amplitude of the expansion is larger in the collective model than in
TDHF. At low temperature,  
the TDHF expansion is more complex than a breathing mode
picture. In this picture, the central density is taken to be the only
relevant variable since a self similar expansion is explicitly assumed. In
fig. \ref{fig6_den}. we show a typical example of density profile for
different times of the evolution for $A=191$. From this picture, it is
clear that hollow structures are created during the evolution and that one
density profile at a given time 
could not be obtained from another one by a simple scaling. The
existence of exotic shapes which are created during the expansion, was already
suggested in several works \cite{bubble}. The main novelty found here as we
will see, is that these structures are consequences of the beating of
intrinsic collective modes\cite{breathing}. In order to get a deeper insight, 
we will focus on the collective
dynamics in the following\footnote{
In the rest of the paper, we will discuss only the soft-EOS since 
the effect we are discussing here are generic and, from the qualitative point of view, do not
depend upon the compressibility of the force used in the simulation (see
fig. \ref{fig4_rho}).}.

\subsection{Collective dynamics}

In this section, we have generalized the analysis performed in ref.\cite
{breathing} to many different initial conditions. In TDHF calculation, no
assumption is made on which collective degrees of freedom could develop
during the evolution.

In order to better understand the complex 
dynamics of the density $\rho (r,t)$, we
have plotted in the fig.\ref{fig9_cont} its Fourier transform $\rho
(r,\omega )$ performed over 1500 fm/c. 
This
corresponds to the expected breathing vibration and indeed the radial
dependence of the density variation $\rho (r,\omega )$ is well fitted by the
usual Tassie transition density\cite{Ring}.

\smallskip In fig.\ref{fig9_cont}, we observe several other collective
higher at higher energies. For $\eta =0.8$, we have many additional
waves at frequencies $17$, $20$, $25$ and around $40MeV$. 
The collective motion located at $2*12.5\simeq 25MeV$
corresponds to the two-phonon excitation of the breathing mode vibration.
The other peaks at $17$, $20$ and around $40MeV$ could be associated to
other collective degrees of freedom.

In order to study either a possible anharmonicity or a coupling between
modes, we have investigated the role of the vibration amplitude
changing the initial conditions. In a pure harmonic picture,
frequencies should remain constant and amplitudes of responses should vary
linearly with the initial perturbation. Focussing on the middle and bottom
part of fig. \ref{fig9_cont}, we see first that the frequencies are almost
constant but the amplitude of the response do not depend linearly on the
initial perturbation. Therefore, the nuclei presents
non-linearities and mode couplings but they are not strong enough to destroy
the collective motions. This means that we are still far from a chaotic
regime.

In this chapter, we have pointed out that not only the breathing mode is
excited in our TDHF calculation but also other modes. These modes are highly
non-local in $r$. This induces a complex dynamics and in particular  
may lead to
the creation of hollow structure. Up to now, many different macroscopic
models\cite{Calor,Papp1,simple} have been used in order 
to extract information about
the break-up of highly excited nuclear systems. In these approaches, the
breathing vibration is taken to be dominant and the expansion is assumed to
be a time-dependent self-similar expansion of the nucleus. However, this
picture fails to reproduce our calculation which indicate that we
must extend our collective phase space to include more degrees of freedom.
In the next chapter, we will address this question and illustrate a possible
generalization the EOS of a finite nucleus in order to include hollow
structures.

\section{Inclusion of Hollow shapes of the
nucleus.}

\begin{figure}[tbph]
\begin{center}
\includegraphics*[height=15cm,width=8cm]{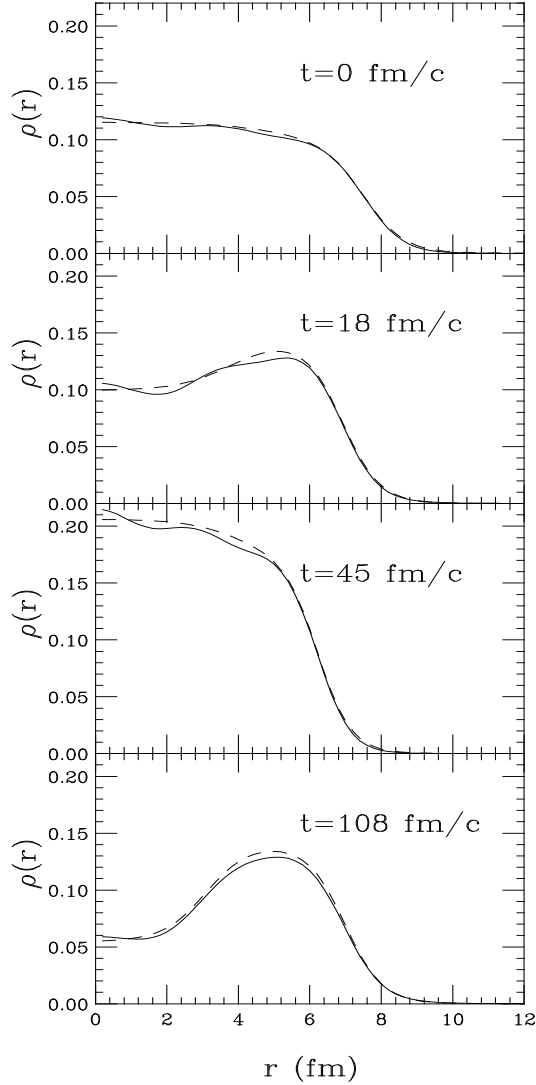}
\end{center}
\caption{Density profiles taken when the central density reaches its
extrema. Figures corresponds to an initial mass $A=191$ and time $%
t=0,18,45 $ and $108fm/c$ from top to bottom, which corresponds to turning
points of the central density. Solid line corresponds to density profiles
obtained in TDHF calculation using a soft EOS. 
Dashed lines corresponds to fit obtained with
two wood-saxon (see text).}
\label{fig6_den}
\end{figure}

\begin{figure}[tbph]
\begin{center}
\includegraphics*[height=15cm,width=10cm]{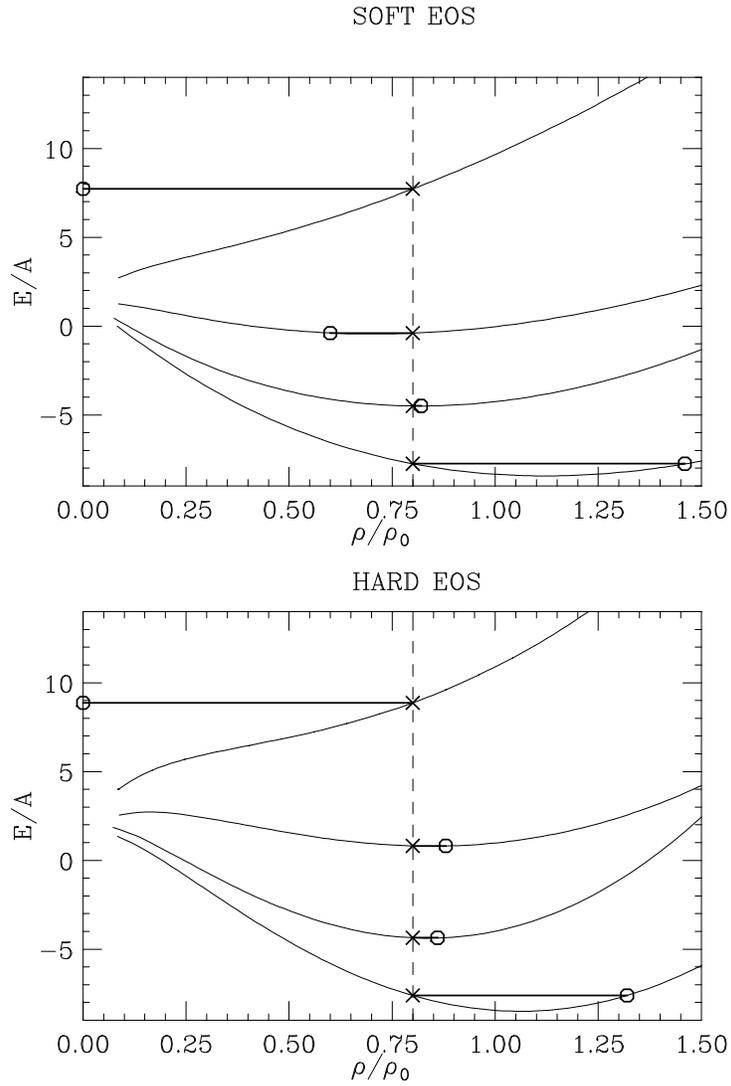}
\end{center}
\caption{Shematic construction of the collective turning point using the
isentropic curve in the excitation energy versus density plot, asking for
both a conservation of the excitation energy and the entropy. Initial
conditions are reported as crosses whereas first turning points are denoted
by circles.}
\label{fig7_sch}
\end{figure}

\begin{figure}[tbph]
\begin{center}
\includegraphics*[height=16cm,width=8cm]{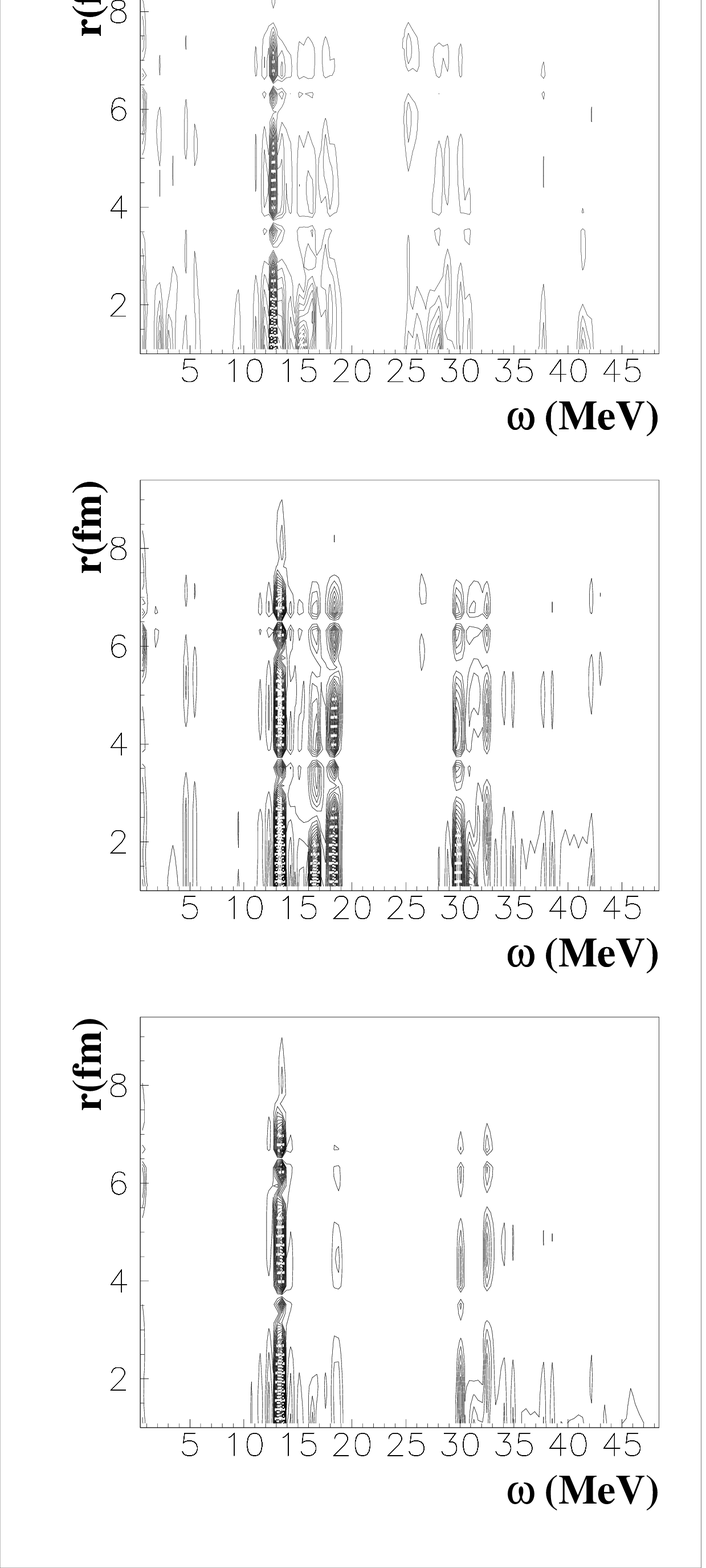}
\end{center}
\caption{Fourier transform $\rho(r,\omega)$ (Left) $\rho(r,t)$. 
From top to bottom three different initial
compression/dilatation factor are considered: $\eta=0.8, 1.0$ and $1.2$
for a nucleus of mass $A=191$.}
\label{fig9_cont}
\end{figure}

During the evolution, we have seen that the nucleus could reach low density
region by creating a hole at its center. In order to study
this type of shape, we have parameterized the density profile as follow: 
\begin{eqnarray}
\rho (r,t)={\huge {\cal A}}\left( \frac{\rho _{1}\cdot \alpha (t)^{3}}{%
1+\exp \left( \frac{r\cdot \alpha (t)-R_{1}}{a_{1}}\right) }+\frac{\rho
_{2}/\beta (t)^{3}}{1+\exp \left( \frac{r/\beta (t)-R_{2}}{a_{2}}\right) }%
\right) 
\label{eq:2WS}
\end{eqnarray}
In this expression, ${\cal A}$ is fixed in order to conserve the number of
particles. The coefficients $\rho _{i}$, $R_{i}$ and $a_{i}$ are fixed by
fitting the density obtained during the evolution of the nucleus $A=191$
at time $t=108fm/c$ (see figure \ref{fig6_den}). $\rho _{1}$, $R_{1}$ and $%
a_{1}$ are respectively equal to $0.0826fm^{-3}$, $3.09fm$ and $0.598fm$
whereas $\rho _{2}$, $R_{2}$ and $a_{2}$ are equal to $0.1354fm^{-3}$, $%
6.98fm$ and $0.528fm$. With this parameterization, the information 
contained in
the density profile is now reduced to only two coefficients $\alpha (t)$ and $%
\beta (t)$. If $\alpha (t)$ is less than zero, the nucleus presents a hole
at its center, otherwise, it presents a bump. Since $\rho _{2}$ is much
bigger than $\rho _{1}$, for low values of $\alpha $, the parameter 
$\beta $ could be approximatively seen
as a global scaling of the nucleus ($\beta >1$ corresponds to a dilatation
whereas $\beta <1$ is a compression). 

In figure \ref{fig6_den}, we have
traced results of the best fit we have obtained for the different density
displayed. Values of $\alpha $ and $\beta $ are reported in table \ref{table5}.

We can see that this parameterization seems to be particularly suitable and
represents well the different density profiles, so that the time
dependent evolution could be accurately replaced by the evolution of the two
parameters $\alpha $ and $\beta $. Since we now include a more general class
of density shapes, this phase-space is a generalized approach of the
breathing mode picture, the two variables being now interpreted as
two collective coordinates.

\subsection{Generalization of the zero temperature equation of state.}

Using the parameterization (\ref{eq:2WS}) we can define the EOS in the $(\alpha
,\beta )$ phase space by computing the energy at zero temperature
for different $\alpha$ and $\beta$ parameters. 
Since eq. (\ref{eq:2WS}) defines only the density profile and not the
wave-functions, we have used the Extended Thomas Fermi formalism
described in \cite{Bra1} where two additional terms are added to the usual
Thomas Fermi kinetic energy
\begin{eqnarray}
E_{K}(r)=E_{TF}(r)+\frac{\hbar ^{2}}{2m}\left( \frac{1}{36}\frac{\left(
\nabla \rho (r)\right) ^{2}}{\rho (r)}+\frac{1}{3}\Delta \rho (r)\right) 
\end{eqnarray}
In this expression, $E_{TF}$ take the form (\ref{eq:EKIN}). The total energy
is calculated by incorporating the potential (\ref{Skyrme}) and integrating
over the r-space.

\begin{figure}[tbph]
\begin{center}
\includegraphics*[height=16cm,width=12cm]{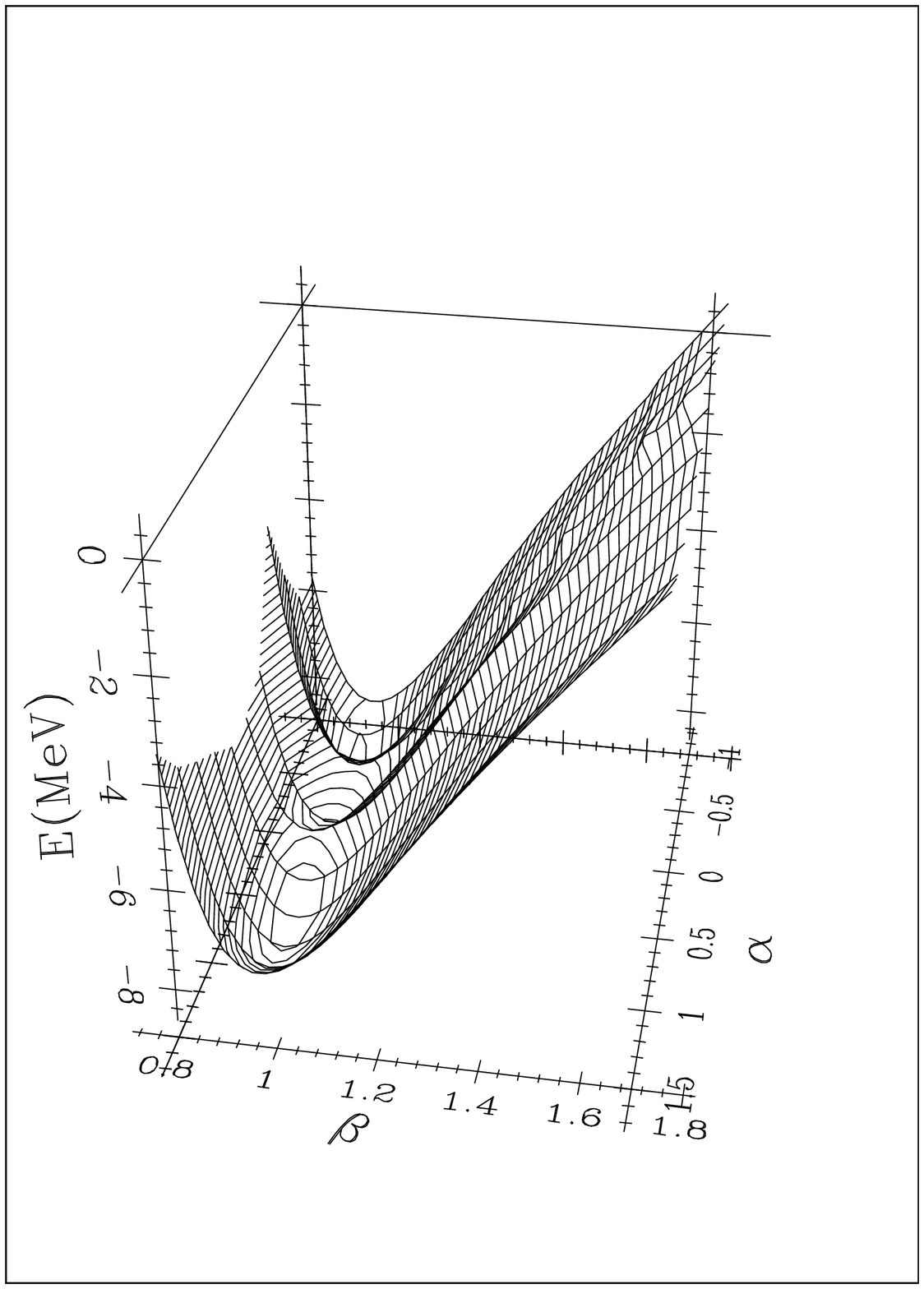}
\end{center}
\caption{Representation of a generalised EOS which include hollow
structures. The energy is reported as a function of $\alpha$ and $\beta$. $%
\alpha$ greater than $0$ corresponds to nucleus with a bump whereas $%
\alpha$ less than one corresponds to a nucleus with a hole at its center.
On the other side, $\beta > 1$ is equivalent to a compression
and $\beta < 1$ means a dilatation.}
\label{fig10_EOS2}   
\end{figure}

By varying the parameters, we can have the generalized EOS: $E=E(\alpha
,\beta )$. This EOS is represented in fig.\ref{fig10_EOS2}. The graphic 
of a self-similar picture displayed in figure \ref{fig7_sch} for $A=191$ 
could be recovered by cutting the graphic with a surface $\alpha = f(\beta)$
(which could be roughly approximated by $\alpha \simeq cte$.
In this picture, we can
see different potential wells with different minima with almost degenerated
energies separated by small barriers (of the order of $1-2$
MeV). These three minima correspond respectively
to a nucleus with a hole ($\alpha <0$), a standard Wood-Saxon shape ($\alpha
=0$) and a nucleus with a bump at the center ($\alpha >0$).

\begin{figure}[tbph]
\begin{center}
\includegraphics*[height=10cm,width=12cm]{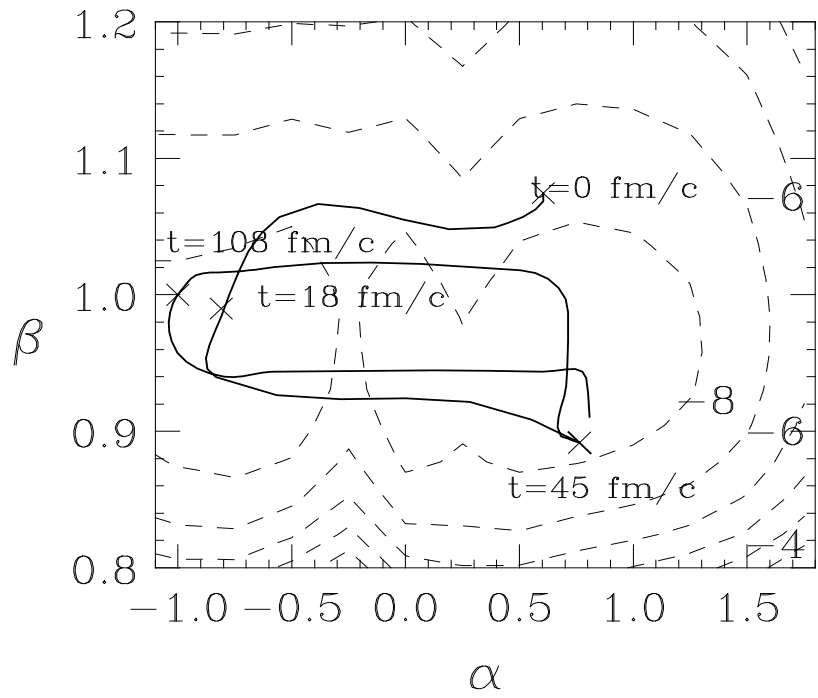}
\end{center}
\caption{Dynamical Path of TDHF simulation in the $(\alpha,\beta)$
phase-space. Contour plot (dashed lines) of the energy are also displayed
(contour corresponds to $-8, -7, -6, -5$ and $-4 MeV$.}
\label{fig11_PATH}
\end{figure}

In order to study the evolution of the system in this $E(\alpha,\beta)$
phase-space, we
have performed a fit of the density profile at each time. Results is
shown up to $108fm/c$ in figure \ref{fig11_PATH}. The path in
this phase-space appears complex and cannot be reduced to a
scaling of the $\beta$ variable. This is another indication of the
fact that we cannot reduce the TDHF dynamics to a breathing mode picture.
During the evolution, we see that the nucleus could go in many different
configurations in this phase space.

In summary, the dynamical expansion as the one studied by 
\cite{Papp1} assumes a predominance
of the breathing mode. In a microscopic calculation, other collective 
vibrations can be excited during the dynamical evolution. This lead to a
complex dynamics for hot and compressed (or dilated) systems for moderate
energies. As 
a direct consequence, the dynamics should be discussed in an enlarged
phase-space of several collective degrees-of-freedom. The illustrative pictures
(\ref{fig10_EOS2} and \ref{fig11_PATH}) seem to indicate that energetically
hollow and uniform nuclei should be both taken into account on the same 
level. Note however, that for initial temperature greater than $5$ MeV,
many orbitals are occupied and levels crossing seems less important and
we do not observe this effect anymore.

Finally, we want to mention that the theoretical 
determination of an effective temperature
during the evolution of the expanding source, is often a hard task and is 
often not possible since the system is only equilibrated at the initial time. 
During the dynamics, the nucleus is strongly out-of equilibrium and
no energy and occupations numbers of levels could be defined on the same 
time (no approximative Fermi-Dirac distribution could be defined
\footnote{Indeed,
during the evolution the Hamiltonian $\hat{h}$ and the density $\hat{\rho}$
do not commute.}).
However, we can imagine another way of defining the temperature. Indeed,
it is possible to follow the density and the energy of the excited system
with time. In a breathing mode picture, we have a univocal correpondance
between these two parameters and the temperature: i.e. the EOS $E=E(\rho,T)$.
However, this picture breaks in our calculation since we have to consider
at least $E=E(\alpha,\beta,T)$ and in fact two parameters are often
not enough. 
In conclusion, the determination of an effective
temperature in models as the one we are considering, is not well-defined
as far as no equilibration process is added on top of the mean-field.   
 
\section{Discussion}

We have used a quantum microscopic simulation in order to discuss the 
interpretation of the caloric curve recently proposed by Papp and 
N\"{o}renberg\cite{Papp1}. In their work the experimentally observed 
caloric curve may be interpreted as a line of turning-point reached  
in the low density region
during the monopole expansion of excited nuclei where
the nucleus could break. Using analogous initial conditions, 
we have pointed out
that  
many differences between a microscopic approach and the collective 
model.
Indeed at low and moderate energies, the TDHF dynamics indicates that 
different collective vibrations could develop on
top of the self-similar vibration. This collective behavior may lead to the
formation of hollow structures. Since the density come from a 
superposition of many modes, we see that the lowest density region could be
reached after a longer time (see fig. \ref{fig4_rho}) than the first
breathing of the nucleus. It should be noticed that experimental DATA of
photon detection seems to indicate such a behavior\cite{Sch1}. The presence
of many collective modes in the dynamic has led us to generalize the simple
breathing mode picture. By a suitable parameterization of the density, we
have defined a generalized Equation Of State. This new approach, include the
possible existence of exotic shape in the dynamic. As we have seen, the
nucleus has a complex path in this new phase-space. On the other
hand,
at high initial excitation energy, when the breathing collective mode
dominates, the amplitude of the monopole oscillation is also different 
from the one
reported in\cite{Papp1}. Since our simulations demonstrate 
that the collective dynamics of an
abraded nucleus could be 
complex and depends on the model used.
The possible interpretation of the caloric curve as a line of turning-point
has to be considered carefully.

\section{Acknowledgements}
We thank J.P. Wieleczko and A. Chbihi 
for helpfull discussion on the manuscript.

\newpage

\newpage

\begin{table}[tbp] \centering%
\begin{tabular}{|l|c|c|}
\hline
\hspace{3cm} & Soft EOS ($SkM^{*}$)   &  Hard EOS ($SIII$) \\ \hline\hline
$\frac{E}{A}$ (MeV) & $-15.77$   &  $-15.85$ \\ 
$\sigma $ & $\frac{1}{6}$   &  $1$ \\ 
$\rho _{0}$ (fm$^{-3}$) & $0.16$   &  $0.145$ \\ 
$t_{0} $ (MeV fm$^3$)& $-2191.73$   &  $-435.41$ \\ 
$t_{3}$ (MeV fm$^{3(1+\sigma)}$)& $18818.8$   &  $17258.8$ \\ 
$a$ (fm)& $0.45979$   &  $0.45979$ \\ 
$V_{0}$ (MeV) & $-461.07$   &  $-401.77$ \\ 
$K_{\infty }$ (MeV) & $198.88$   &  $368.4$ \\ \hline
\end{tabular}
\caption{Parameters and properties of the two interactions
used in this article.\label{table1}}%
\end{table}

\newpage

\smallskip

\begin{table}[tbp]
\begin{center}
\begin{tabular}{|c|c|c|}
\hline
\ $A_{i}$ & $S_{i}$ Soft EOS ($SkM^{*}$)   &  $S_{i}$ Hard EOS ($SIII$) \\ 
\hline\hline
191 & 0.44   &  0.37 \\ 
187 & 0.38   &  0.35 \\ 
183 & 0.45   &  0.51 \\ 
177 & 0.63   &  0.51 \\ 
172 & 0.79   &  0.71 \\ 
165 & 0.97   &  0.92 \\ 
158 & 1.14   &  1.12 \\ 
150 & 1.33   &  1.31 \\ 
142 & 1.52   &  1.50 \\ 
134 & 1.59   &  1.55 \\ 
120 & 1.80   &  1.82 \\ 
111 & 2.0   &  2.06 \\ 
98  & 2.46   &  2.49 \\ 
89 & 2.79   &  2.80 \\ \hline
\end{tabular}
\end{center}
\caption{Examples of initial conditions considered. For each initial 
masses and for the two considered forces, initial entropies are given.}
\label{table2}
\end{table}

\newpage

\begin{table}[tbp] \centering%
\begin{tabular}{|c|c|c|}
\hline
\ $A_{i}$ \ \ \ \ \ \  & $\rho _{0}\left( A_{i}\right) $ Soft EOS ($SkM^{*}$) 
  &  $\rho _{0}\left( A_{i}\right) $ Hard EOS ($SIII$) \\ \hline\hline
191 & 0.142   &  0.139 \\ 
187 & 0.142   &  0.139 \\ 
183 & 0.143   &  0.140 \\ 
177 & 0.147   &  0.142 \\ 
172 & 0.150   &  0.144 \\ 
165 & 0.155   &  0.147 \\ 
158 & 0.160   &  0.150 \\ 
150 & 0.165   &  0.153 \\ 
142 & 0.170   &  0.157 \\ 
134 & 0.155   &  0.145 \\ 
120 & 0.143   &  0.140 \\ 
111 & 0.143   &  0.140 \\ 
98  & 0.154   &  0.148 \\ 
89 & 0.163   &  0.154 \\ \hline
\end{tabular}
\caption{Saturation densities for the various initial
 masses.\label{table3}}%
\end{table}%

\newpage

\newpage

\newpage

\newpage

\begin{table}[tbp] \centering%
\begin{tabular}{ccc}
 & Soft EOS ($SkM^{*}$)  &   Hard EOS ($SIII$) \\ 
\begin{tabular}{|c|}
\hline
\ $A_{i}$ \ \ \ \ \  \\
\\ \hline\hline
191 \   \\ 
150 \  \\ 
111 \   \\ 
89 \\ \hline
\end{tabular}
& 
\begin{tabular}{|c|}
\hline
$\rho _{\min }/{\rho_0}$ \\
from [4] \\ \hline\hline
0.68 \\ 
0.4 \\ 
0.0 \\ 
0.0 \\ \hline
\end{tabular}
\begin{tabular}{|c|}
\hline
$\rho _{\min }/\rho_0$ \\
schem \\ \hline\hline
1.45 \\ 
0.82 \\ 
0.6 \\ 
0.0 \\ \hline
\end{tabular}
\begin{tabular}{|c|}
\hline
$\rho _{\min }/\rho_0$ \\
exact \\ \hline\hline
0.408 \\ 
0.764 \\ 
0.77 \\ 
0.25 \\ \hline
\end{tabular}
&
\begin{tabular}{|c|}
\hline
$\rho _{\min }/{\rho_0}$ \\
from [4] \\ \hline\hline
0.75 \\ 
0.52 \\ 
0.23 \\ 
0.0 \\ \hline
\end{tabular}
\begin{tabular}{|c|}
\hline
$\rho _{\min }/\rho_0$ \\
schem \\ \hline\hline
1.32 \\ 
0.86 \\ 
0.89 \\ 
0.0 \\ \hline
\end{tabular}
\begin{tabular}{|c|}
\hline
$\rho _{\min }/\rho_0$ \\
exact \\ \hline\hline
0.67 \\ 
0.80 \\ 
0.786 \\ 
0.494 \\ \hline
\end{tabular}
\end{tabular}
\caption{Comparison between the turning point observed dynamically and
those predicted using a schematic self-similar isentropic 
dynamics at constant energy (see figure 6). Turning point given
in [4] for the same forces are also reported.
\label{table4}}%
\end{table}%

\newpage

\newpage

\smallskip 

\newpage

\newpage

\begin{table}[tbp] \centering%
\begin{tabular}{|r|r|r|}
\hline
$t$ (fm/c) & $\alpha (t)$ & $\beta (t)$ \\ \hline
0 & 0.60 & 1.07 \\ \hline
18 & -0.80 & 0.98 \\ \hline
45 & 0.76 & 0.89 \\ \hline
108 & -1.00 & 1.00 \\ \hline
\end{tabular}
\caption{Different values of $\alpha (t)$ and $\beta (t)$ obtained by fitting 
the density profile at divers time of the expansion of the nucleus $A=191$.
\label{table5}}%
\end{table}%

\newpage

\newpage

\end{document}